\documentclass[prl,preprint,superscriptaddress,amsmath,amssymb,floatfix,showpacs]{revtex4}
\usepackage{graphicx}
\usepackage{epsfig}
\usepackage{dcolumn}
\usepackage{bm}
\usepackage{psfrag}
\begin{document}

\bibliographystyle{apsrev}

\title{Complete Wetting of Nanosculptured Substrates}

\author{M. Tasinkevych}
\affiliation{Max-Planck-Institut f\"{u}r Metallforschung,
             Heisenbergstr. 3, D-70569 Stuttgart, Germany}
\affiliation{Institut f\"{u}r Theoretische und Angewandte Physik, 
         Universit\"{a}t Stuttgart, Pfaffenwaldring 57, 
         D-70569 Stuttgart, Germany}
\author{S. Dietrich}
\affiliation{Max-Planck-Institut f\"{u}r Metallforschung,
             Heisenbergstr. 3, D-70569 Stuttgart, Germany}
\affiliation{Institut f\"{u}r Theoretische und Angewandte Physik, 
         Universit\"{a}t Stuttgart, Pfaffenwaldring 57, 
         D-70569 Stuttgart, Germany}

\date{\today}

\begin{abstract}
Complete wetting of geometrically structured substrates by
one-component  fluids with long-ranged interactions is studied. We consider
periodic arrays of rectangular or parabolic grooves and lattices
of cylindrical or parabolic pits. We show that the midpoint interfacial
heights within grooves and pits are related in the same way as for complete
wedge and cone  filling. For sufficiently deep cavities with vertical walls
and small undersaturation, an effective planar scaling regime emerges.
The scaling exponent is $-1/3$ in all cases studied, and only the amplitudes
depend on the geometrical features. We find quantitative agreement with recent
experimental data for such systems.

\end{abstract}

\maketitle


The growing interest in device miniaturization has led to the emergence of various experimental
techniques  of tailoring  the geometrical and chemical topography of solid surfaces  at  mesoscopic scales \cite{fabrication,lotus_eff}. 
Such nano-patterning of the surfaces  may result in  drastic changes of their wetting characteristics, which
is important for technologies such as  micro-fluidics \cite{microfluidic} or the fabrication
of  super-hydrophobic or super-hydrophilic surfaces  \cite{lotus_eff}.  
Experimental studies of  complete wetting on sculptured surfaces \cite{gang-exper_prl:05,bruschi-exper_prl:02}
demonstrate the strong influence of nanocavities on the  adsorption behavior relative to that of flat substrates.
Theoretical studies of  adsorption in  infinitely deep generalized wedges \cite{parry_nature:00}
predict  geometry dependent wetting exponents.

Although  it is  known that a non-planar topography of a substrate modifies  its wetting by a fluid, 
recent  studies  have revealed surprising  hidden symmetries, or so-called covariances,
which  relate various local adsorption properties
for different substrate geometries. These covariances imply that different confining potentials can lead 
to identical local interfacial properties once external fields
 are suitably rescaled.
\begin{figure}[]
\begin{center}
\includegraphics[width=5cm]{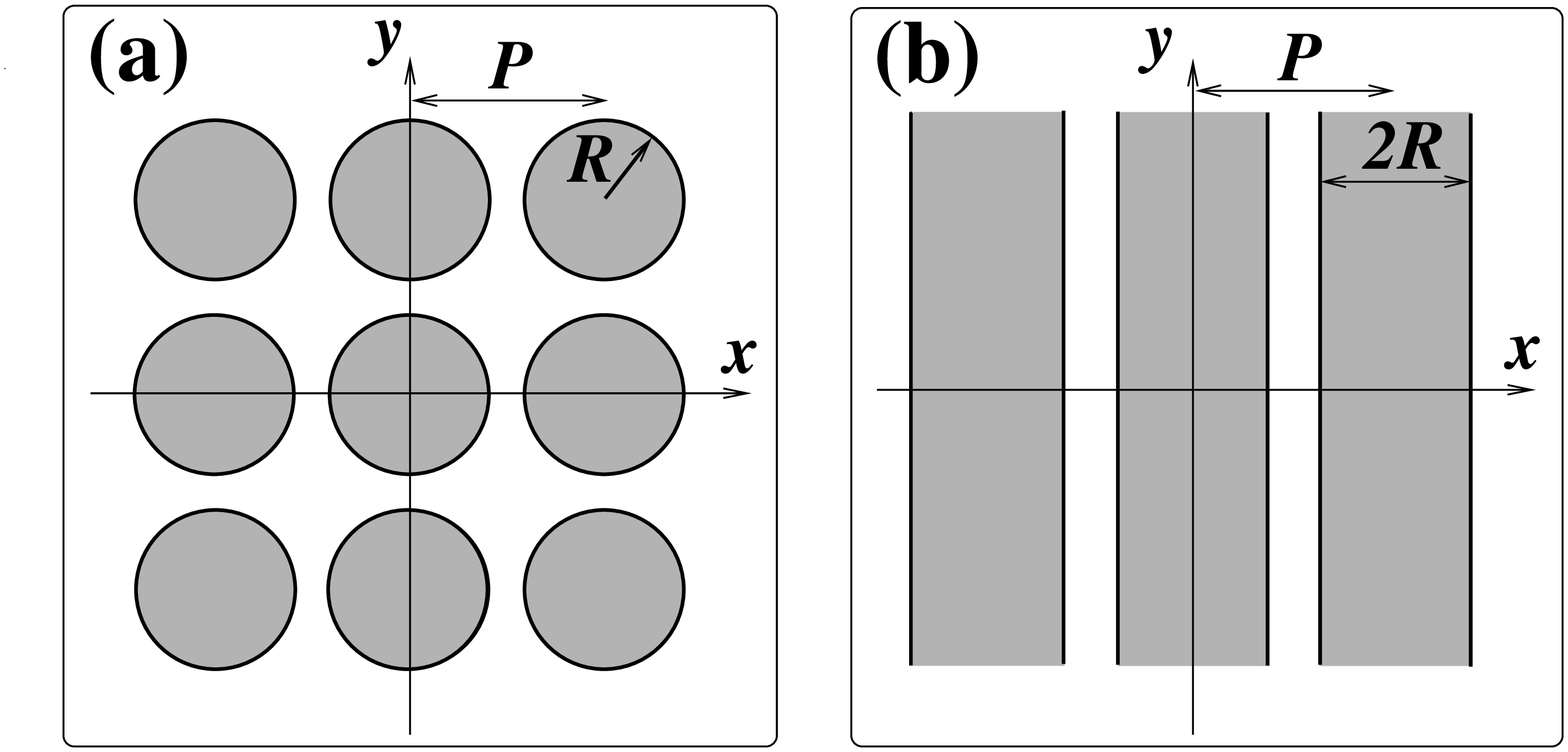}
\end{center}
\caption{ Top view of the substrate geometries. (a) Quadratic lattice 
 of  identical  pits (cylinders or paraboloids).
(b) Periodic array  of  grooves (rectangular or parabolic). 
All cavities have finite depths $D$.
}
\label{geometry}
\end{figure}
%
For instance, for both long- and short-ranged forces, complete wetting at the apex of a wedge can be
 mapped onto  critical  wetting of a planar substrate  
with the apex angle playing the role of the contact angle \cite{parry_prl:03}, which maps one-to-one to temperature.
  In two-dimensional systems,  critical wedge filling  can be related  to  the strong-fluctuation regime of critical planar wetting
\cite{abraham_epl:02}. Recently, a new  example of geometrical covariance relating  wedge and
cone complete filling  has been reported  \cite{parry_prl:05} showing that the equilibrium midpoint
interfacial heights  $l^{(0)}$ in a {\it c}one and a {\it w}edge obey the relation $l_c^{(0)}(\Delta\mu,\alpha) = l_w^{(0)}(\Delta\mu /2,\alpha)$ with
$\alpha$ being the substrate tilt angle and $\Delta\mu\geq 0$  the chemical potential deviation from liquid-vapor coexistence.
 This relation is valid for the leading behaviors of $l$ in the limit 
 $\Delta\mu \rightarrow 0^+$.

Based on an effective interface Hamiltonian in the following we 
 demonstrate that  complete wetting of substrates patterned by  periodic  arrays of grooves 
or quadratic lattices  of pits (see Fig.~\ref{geometry}),  both  of depths $D$,  exhibits a
 geometrical covariance similar to the one described in Ref.~\cite{parry_prl:05}.
We consider rectangular or parabolic  grooves and cylindrical or parabolic pits, taking into account the  long
range of the intermolecular 
potentials. 
We observe four different scaling regimes: filling, post-filling, effective planar, and planar,
with the neighboring regimes being separated by, as we call, $\Delta\mu_{fil}^{p,g}>\Delta\mu_{\pi}^{e}>\Delta\mu_{\pi}$.
The aforementioned covariance relates the behavior of the midpoint wetting film thicknesses  for all
geometries and holds within an undersaturation range $\Delta\mu_{\pi}^{e}\lesssim\Delta\mu\lesssim\Delta\mu_{fil}^{p,g}(R)$
(for $\Delta\mu_{\pi}^{e}$ see below, the superscripts $p,g$ refer to {\it p}its and {\it g}rooves, respectively) which we call
the  {\em post-filling scaling regime}.
 In the case of cylindrical pits or rectangular grooves, for $\Delta\mu\searrow \Delta\mu_{fil}^{p,g}(R)$ 
and for sufficiently large $D/R$, the analogue of  capillary condensation occurs such that the pits or grooves are
 rapidly, but continuously, filled by the liquid.  However, in the case of the parabolically shaped pits or grooves,
$\Delta\mu_{fil}^{p,g}$ marks the crossover from the continuous power-law filling regime at $\Delta\mu>\Delta\mu_{fil}^{p,g}$, to the   
post-filling scaling regime. 
We find numerically for  $R/\sigma\gtrsim 50$ (see Fig.~\ref{geometry}), where $\sigma$ is a molecular length scale,  
 $\Delta\mu_{fil}^{p,g}(R)\sim R^{-1-\delta}$ with a small positive effective
exponent $\delta$, and $\Delta\mu_{fil}^{p} =2\Delta\mu_{fil}^{g}$ . If we denote the equilibrium interface height
at the position of the symmetry axes of the cylindrical or parabolic pits as $l_p^{(0)}$, and in the middle of the rectangular or parabolic
 grooves as $l_g^{(0)}$,  we obtain in the post-filling regime
\begin{eqnarray}
l_{p,g}^{(0)}(\Delta\mu,R,P,D)&=&R\Lambda_{p,g}\Bigl ( (\Delta\mu/\varepsilon_f) (R/\sigma)^{1+\delta} \Bigr ), \nonumber \\
\Lambda_p(x)&=&\Lambda_g(x/2).
\label{eq:main_1}
\end{eqnarray}
 The scaling functions 
$\Lambda_{p,g}(x)$ do not depend on  $D$  and $P$, i.e., in this regime the midpoint interfacial heights 
increase upon decreasing undersaturation
in the same way for an isolated cavity as for arrays of them; $\varepsilon_f$ is a molecular energy scale.

For $\Delta\mu<\Delta\mu_{\pi}^{e}$ the  cavities are completely filled by the liquid and the equilibrium interface 
heights $l_{p,g}(x,y)=l_{p,g}$  become de facto independent of the lateral coordinates ${\bf x}\equiv (x,y)$. 
In the case of the rectangular grooves or cylindrical pits, 
$\Delta\mu_{\pi}^{e}$ marks the crossover from the post-filling scaling regime
to the {\em effective planar scaling regime} within which the wetting behavior of geometrically patterned substrates can be  mapped 
onto that of  layered flat solids.
 The upper layer of those solids has a thickness $D$, and its composition is related 
to the geometrical parameters  $R$ and $P$. This results in the following scaling relations for the
effective planar scaling regime:
\begin{eqnarray}
l_{p,g}(\Delta\mu,R,P,D)=(\Phi_{p,g})^{\frac{1}{3}}l_{\pi}(\Delta\mu),
\label{eq:gc_planar_H}
\end{eqnarray} 
where $l_{\pi}$ is the thickness of the adsorbed liquid film on the corresponding non-patterned planar substrate. 
$\Phi_{p}=1-\pi(R/P)^2$ and $\Phi_{g}=1-2R/P$ are the areal fractions of solid in the top layer, $z=0$, of substrates
with cylindrical pit and rectangular groove patterns, respectively.
This result is reminiscent
of the Cassie equation \cite{cassie_48} describing the apparent contact angle on  chemically structured
substrates. For $D\rightarrow 0$ the width of the effective planar regime, i.e., the range 
of applicability of Eq.~\ref{eq:gc_planar_H} vanishes. 


 
 Finally, at $\Delta\mu=\Delta\mu_{\pi}\sim D^{-3}$ with $\Delta\mu_{\pi}<\Delta\mu_{\pi}^{e}$ the systems 
cross over to the {\em planar scaling} regime, in which the geometrical patterns are irrelevant.
This crossover occurs only for  long-ranged dispersion forces; we consider only them throughout. (For  short-ranged interactions, instead,
the  growth of the wetting film would remain  determined by the areal fraction of solid at $z=0$ for all film thicknesses.) 
In the case of the parabolic pits and grooves, we do not observe the effective planar scaling regime.  There is rather 
an extended crossover region from the post-filling scaling regime to the planar one.

\begin{figure}[]
\begin{center}
\psfrag{go}{${\tiny \pmb{\rightarrow}}$}
\psfrag{infty}{${\tiny \pmb{\infty}}$}
\includegraphics[width=6.5cm]{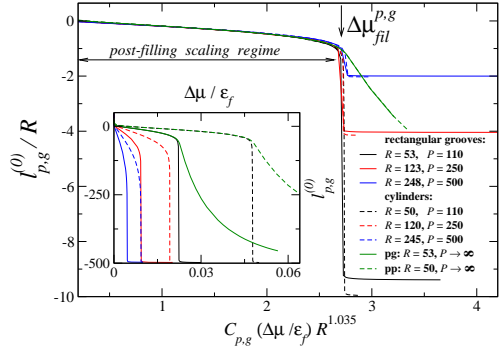}
\end{center}
\caption{  Interfacial heights $l_{p,g}^{(0)}$ (measured from the top substrate layer $z=0$) at the middle of the
cylindrical and  parabolic pits, $l_p^{(0)}$ (dashed lines), and of the
rectangular and parabolic  grooves, $l_g^{(0)}$  (solid lines), as a function of 
 $\Delta\mu$
with $C_p=1$ and $C_g=2$. Rescaling the variables  according to Eq.(\ref{eq:main_1}) leads to data collapse. 
The inset shows the same data unscaled.  The depth of the cavities is $D=500$; $pp(g)$ denotes
parabolic pits (grooves). All length are measured in units of $\sigma$. 
}
\label{fig2}
\end{figure}

We base our calculations on the effective interface Hamiltonian description \cite{dietrichbook}
\begin{equation}
H[l] = \iint d^2x \Bigl  (\sigma_{lg}\sqrt{1+(\nabla l)^2} +
 \Delta{\mu} \Delta\rho l + W({\bf x},l) \Bigr ),
\label{eq:hamiltonian}
\end{equation}

\noindent where $l({\bf x})$ denotes the local interfacial height which is measured from  the plane $z=0$ where the substrate ends, 
$\sigma_{lg}$ is the surface tension of the free liquid-vapor interface,
$\Delta\rho=\rho_l-\rho_g$ is the difference in number densities of the coexisting bulk phases, 
and $W({\bf x},l)$ is the effective interface potential. The
functional in Eq.~(\ref{eq:hamiltonian}) can be derived systematically from microscopic density functional theory, 
which allows one to determine the explicit functional form of $W({\bf x},l)$ for a given substrate shape and
its dependence on the {\it f}luid-fluid and {\it s}ubstrate-fluid interaction potentials.
 We approximate the attractive parts of the pair potentials by 
$\phi_{f,s}(r)  = -\frac{128\sqrt{2}}{9\pi}\varepsilon_{f,s}\sigma^6(\sigma^2+r^2)^{-3}$.
The amplitude is chosen such that the integrated strength of $\phi_f$ equals that of the 
attractive contribution of the  Lennard-Jones potential obtained by a strict application of the 
Weeks-Chandler-Andersen procedure.
 This leads to the effective interface potential 
\begin{eqnarray}
W_{\Omega}({\bf x},l)&=&A\times I_{\Omega}({\bf x},l), \nonumber \\
 I_{\Omega}({\bf x},l) &=& \int_{l}^{\infty}dz\int_{\Omega}d^3r^{\prime}(\sigma^2+|{\bf r}-{\bf r^{\prime}}|^2)^{-3},
\label{eq:bind_potential}
\end{eqnarray}
 where $A=-\frac{128\sqrt{2}}{9\pi}\sigma^6\Delta\rho(\rho_l\varepsilon_{f}-\rho_s\varepsilon_{s})>0$ is
an effective  Hamaker constant with $\rho_s$ as the number density of the substrate, and $\Omega$ denotes the 
domain occupied by  substrate particles.  
The effective interface potential of  the planar substrate is
$W_{\pi}(l\gg\sigma)\approx A\pi/(12l^2)$ so that $l_{\pi}(\Delta\mu\rightarrow 0)=(A\pi/(12\Delta\rho\Delta\mu))^{1/3}$.

In the following we minimize the functional in Eq.~($\ref{eq:hamiltonian}$) numerically, which yields the equilibrium interface height
$l({\bf x})$ within mean-field theory which is valid in $d=3$ for complete wetting
\cite{dietrichbook} and filling \cite{parry_prl:05} for the dispersion forces considered here.
 For all substrate geometries the midpoint heights $l_{p,g}^{(0)}(\Delta\mu)$ exhibit four different regimes.
The first regime corresponds to the filling of the cavities. For the case of rectangular grooves and cylindrical pits,
and for $D/R$ large enough, a quasi abrupt, but still continuous filling of the  cavities takes place at $\Delta\mu=\Delta\mu_{fil}^{p,g}$, 
which is shown in the inset of Fig.~{\ref{fig2}}. A similar behavior has been reported earlier for an isolated  rectangular groove \cite{darbellay}.
We find that the locus  $\Delta\mu_{fil}^{p,g}$ of the filling transformation scales with the  
lateral size of the  cavities as $R^{-1-\delta}$, with an effective exponent $\delta\approx0.035$ for both 
the rectangular grooves and the cylindrical pits.
For the case of parabolic pits and grooves, complete filling of the structures is described by an
effective power law, $l_{p,g}^{(0)}(\Delta\mu,R,D)\sim\Delta\mu^{-\gamma(R,D)}$  valid for $\Delta\mu \gtrsim \Delta\mu_{fil}^{p,g}$. 
We find  the values of the effective exponent $\gamma$ ranging from ca. $ 3.1$, for $R=245\sigma$, to ca. $ 2.0$, for $R=50\sigma$, 
at a cavity depth $D=500\sigma$. Moreover, the complete filling of the parabolic cavities obeys the covariance relation
$l_p^{(0)}(\Delta\mu,R,D)=l_g^{(0)}(\Delta\mu/2,R,D)$.

In the second, post-filling regime, i.e., for $\Delta\mu\in (\Delta\mu_{\pi}^e,\Delta\mu_{fil}^{p,g})$, the midpoint height
for all  patterns shows an almost linear dependence on $\Delta\mu$ on normal scales. The morphologies of the liquid films still 
reflect the geometrical patterns, i.e., there are  
considerable lateral variations of the interfacial heights.  
We find that the slopes of the $l_{p,g}^{(0)}$ curves scale as $R^\alpha$ with
$\alpha\approx 2$. Thus, combining  this fact with the scaling of the filling chemical potential $\Delta\mu_{fil}^{p,g}$,
we propose for the functions $l_{p,g}^{(0)}$ in the post-filling regime the scaling forms  given by Eq.~(\ref{eq:main_1}).
The scaling functions $\Lambda_{p,g}$ and the corresponding data collapse upon suitably rescaling
 the chemical  potential are shown in  Fig.~\ref{fig2}.  
The scaling functions  $\Lambda_{p,g}$ in the post-filling regime do not depend on the cavity depth $D$ and the pattern
periodicity $P$. In the case of a single cavity, i.e., in the limit $P\rightarrow\infty$, we obtain 
 the same curves $l_{p,g}^{(0)}(\Delta\mu)$  as those  presented in  Fig.~\ref{fig2}. 

\begin{figure}[]
\begin{center}
\includegraphics[width=6.5cm]{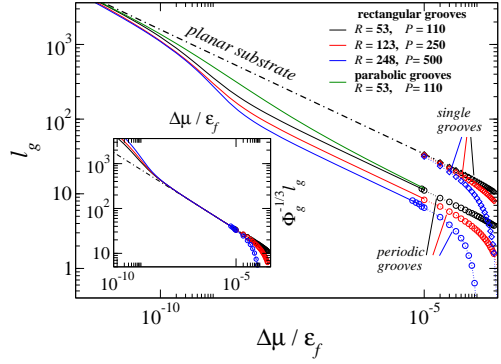}
\end{center}
\caption{Wetting film thickness $l_g$ above  grooves  for small $\Delta\mu$.
 The symbols  represent the midpoint interfacial height $l_g^{(0)}$.
The solid lines are obtained by minimizing Eq.(\ref{eq:hamiltonian}) assuming 
 $l({\bf x })=l_g$. 
On the present scales they are indistinguishable from those for film thicknesses $l_{\pi}^e(\Delta\mu)$ 
on  layered and flat ersatz substrates, with a top layer of height $D$ and
an effective Hamaker constant $A^e=A\Phi_g$, where $A$ is the  Hamaker constant of the solid without grooves.
In the inset the vertical axis is rescaled according to  Eq.~(\ref{eq:gc_planar_H})
leading to data collapse within an intermediate regime, the width of which increases as 
 $\Delta\mu_{\pi}^e-const\times D^{-3}$ for $D\rightarrow\infty$.
The  groove depth is $D=500$.}
\label{fig4}
\end{figure}

Below a certain value $\Delta\mu_{\pi}^e$ the interface height becomes de facto flat, denoted as $l_{p,g}$ for the respective patterns.
In this case the functional in Eq.~(\ref{eq:hamiltonian}) reduces to a function of $l_{p,g}$  
the minimum of which determines the thicknesses of the wetting films in this 
regime. 
Figure~\ref{fig4}  shows the functions $l_g(\Delta\mu)$  for several
values of $R$ and $P$ as well as the planar film thickness $l_{\pi}(\Delta\mu)$.  
The rectangular groove (and cylindrical pit, not shown) geometries  lead to the same power law behavior as for a flat surface, 
$l_{p,g}\sim\Delta\mu^{-1/3}$, but with different amplitudes reflecting
different effective Hamaker constants, which  depend on the geometry of the patterns. 
We call this behavior the effective planar scaling regime.
 $l_{g}(\Delta\mu)$ for the parabolic grooves does not reveal the effective planar scaling regime, but
gradually crosses over to the  planar one, $l_\pi$.
In order to find the geometrical dependence of the amplitudes of the  scaling laws given above,  we mimic sculptured substrates
by layered and flat ersatz solids, with the effective interface potential 
$W(l\gg\sigma)\approx\frac{\pi}{12}\bigl (A^e/l^2+(A-A^e)/(l+D)^2\bigr )$
\cite{robbins}. The first term  is the effective interface potential of a flat semi-infinite solid with Hamaker constant $A^e$ and the
second term is the correction due to the actual bottom part of the substrate, $z\in(-\infty,-D)$, with Hamaker constant $A$.
For $\sigma\ll l\lesssim D$, to a good approximation  one may ignore the bottom part of the substrate 
(i.e., the second term in $W(l\gg\sigma)$) so that the
 amplitude is determined by the Hamaker constant $A^{e}$.
Consider a lateral unit cell $\omega_0$ of this sculptured part of the substrate ($z\in[-D,0]$), with $\omega_s$ as its domain occupied by the solid.
By requiring  $W_{\cup \omega_s}({\bf x},l)=W^{e}_{\cup \omega_0}({\bf x},l)$, we obtain for the effective Hamaker constant
$A^{e}=AI_{\cup \omega_s}({\bf x},l)/I_{\cup \omega_0}({\bf x},l)$, where $W^{e}_{\cup \omega_0}$ is the effective interface
 potential generated by the laterally homogeneous top layer of the layered and flat ersatz substrate;  the symbol  $\cup$ denotes the
union of domains. For $l\gg\sigma$ the integrals $I_{\cup \omega_i}$ ($i=s,0$) can be approximated as:
\begin{equation}
I_{\cup \omega_i} \approx \int_{l}^{\infty}dz\int_{\cup \omega_i}\frac{d^3r^{\prime}}{(z-z^{\prime})^6}
\biggl (1-3\frac{\sigma^2+||{\bf x}-{\bf x}^{\prime}||^2}{(z-z^{\prime})^2 } \biggr). 
\label{eq:int_ratio}
 \end{equation}

\begin{figure}[]
\begin{center}
\includegraphics[width=6.5cm]{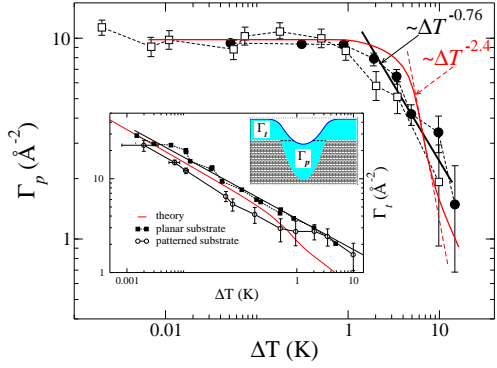}
%
\end{center}
\caption{Liquid adsorption at parabolic pits.  
Inset: $\Gamma_t(\Delta T)$.
Red line -- theory; symbols and black lines -- experimental data from Ref.~\cite{gang-exper_prl:05}
(where $\Gamma_p$ is denoted as $\Gamma_c$).
$\Delta T$ is proportional to $\Delta\mu$ (see main text).}
\label{experiment}
\end{figure}

\noindent In  Eq.~(\ref{eq:int_ratio}) the leading term as function of $l$  renders $A^{e}\approx A S_s/S_0$, where $S_i$ is the surface area
of the domain $\omega_i\cap\{({\bf x},z=0)\}$. Thus, we obtain $A^{e}_g\approx A \Phi_g$ for the rectangular grooves 
and  $A^{e}_p\approx A \Phi_p$
for  cylindrical pits (see Eq.~(\ref{eq:gc_planar_H})). 
The thicknesses  $l_{\pi}^{e}(\Delta\mu)$ of the wetting films   
on such layered ersatz substrates, which effectively correspond to  arrays of grooves, are 
 almost indistinguishable from the corresponding ones for  $l_g$. 
Therefore, we conclude that, for sufficiently  thick wetting films, $l_{g}$ obeys
the scaling relation given in Eq.~(\ref{eq:gc_planar_H}).  Indeed, after rescaling
by the geometry-dependent factors $(\Phi_{g})^{-1/3}$, the curves for $l_{g}$ collapse and, within the numerical precision, 
coincide with the  curve for $l_{\pi}$. This is shown in the inset  of  Fig.~\ref{fig4}. 
In the case of a lattice of cylindrical pits we observe the same behavior of the corresponding $l_{p}$ curves (not shown).
 As  $l_{g}$  reaches the value $\sim D$, i.e., at $\Delta\mu=\Delta\mu_{\pi}\sim D^{-3}$,  
a crossover to the {\it planar scaling regime} takes place. In the planar scaling 
regime $\Delta\mu\lesssim\Delta\mu_{\pi}$ the geometrical patterns are irrelevant. As $D$ decreases,  $\Delta\mu_{\pi}$ eventually merges  
 with  $\Delta\mu_{\pi}^{e}$, eliminating the effective planar scaling regime.   

Figure~\ref{experiment} compares our results with the corresponding experimental data of Ref.~\cite{gang-exper_prl:05}, in which
the adsorption of methylcyclohexane (MCH) on a silicon substrate sculptured by a hexagonal lattice of
parabolic pits has been studied with the following  
 geometrical parameters:  depth  $D\approx 200$\AA;
 radius of the pits at the opening $R\approx 123$\AA; lattice constant $P\approx 394$\AA.  
As in  Ref.~\cite{gang-exper_prl:05}, we display the volume $\Gamma=\Gamma_p+\Gamma_t$
of adsorbed liquid divided by the projected area and multiplied by the electron density of
the bulk fluid as the sum of the amount $\Gamma_p$ adsorbed in the pit, and the amount $\Gamma_t$ 
adsorbed above the pit opening. For MCH we adopt the Lennard-Jones parameters  
$\sigma=5.511${\AA}
and $\varepsilon_f/k_B=446 K$ \cite{goldman}. Then the relation between the undersaturation $\Delta\mu$
and the reservoir-substrate temperature difference (used in the experiment as a means to tune the
deviation from liquid-vapor coexistence) is $\Delta T\approx 32.37 \frac{\Delta\mu}{\varepsilon_f} K$.
Fitting the film thickness on the planar substrate to the corresponding experimental curve fixes the
value of the Hamaker constant. For the liquid-vapor surface tension of MCH
we use  $\sigma_{lg}=22.72\frac{dyn}{cm}$ for  $T=30^{\circ}C$   \cite{jasper}. 
Adjusting the geometry of the patterns to the experimental
ones we calculate $\Gamma_p$ and $\Gamma_t$ as functions of $\Delta T$. 
We  emphasize that at this stage there are no free parameters left in the 
model. The resulting curves are shown in Fig.~\ref{experiment}. For both quantities 
we obtain good quantitative agreement with the experimental results.
In the filling regime $(5K \lesssim\Delta T\lesssim 8K)$  the calculated $\Gamma_p(\Delta T)$ 
exhibits a power-law behavior $\Gamma_p\propto\Delta T^{-\beta_p}$  with an effective exponent 
$\beta_p\approx 2.4$. The apparent disagreement with the value $\beta_p\approx 0.76$ used in Ref.~\cite{gang-exper_prl:05}
can be explained by the mislocation of the filling regime on the experimental $\Gamma_p(\Delta T)$ curve due
to the large error bars for the data points at large $\Delta T$.
The weak crossover at $\Delta T\approx 8K$ corresponds to the disappearance of the interfacial meniscus, 
so that for  larger values of $\Delta T$ the interface follows the shape of the substrate. We have found
that for larger cavities this crossover becomes more pronounced and for sufficiently large  $D$ and $R$ 
we obtain the  exponent $\beta_p = 3.4$.

In summary, as function of undersaturation we have studied  complete wetting  of four classes of substrates structured
by one- and  two-dimensional periodic patterns of  fixed depth and we  have identified four scaling regimes.
The filling and the post-filling evolutions of the interfacial profiles do not depend on the periodicity of the patterns,
but are determined  by a single isolated cavity. For sufficiently  deep  structures there exists a range of undersaturations,
 in which the midpoint interfacial heights, obtained for different patterns, can be expressed in terms of a single scaling function.
For small undersaturations, the single-cavity behavior crosses over to one dominated by the presence of many of them,
for which the interfacial thickness increases as on a planar substrate, but characterized by an effective  geometry-dependent Hamaker constant.
Ultimately, for very small undersaturations the wetting film thickness becomes independent of the 
geometrical substrate structures.



\begin{thebibliography}{99}

\bibitem{fabrication}
T. Yanagishita, K. Nishio, and H. Masuda,
Adv. Mater. {\bf 17}, 2241 (2005).
%
\bibitem{lotus_eff}
E. Martines, K. Seunarine, H. Morgan, N. Gadegaard, C. D. W. Wilkinson, and M. O. Riehle,
Nano Lett. {\bf 5}, 2097 (2005).

\bibitem{microfluidic}
E. Delamarche, D. Junker, and H. Schmid, Adv. Mater. {\bf 17}, 2911 (2005);
and references therein.



\bibitem{gang-exper_prl:05}
O. Gang, K. J. Alvine, M. Fukuto, P. S. Pershan, C. T. Black, and B. M. Ocko,
Phys. Rev. Lett. {\bf 95}, 217801 (2005).

\bibitem{bruschi-exper_prl:02}
L. Bruschi, A. Carlin, and G. Mistura,
Phys. Rev. Lett. {\bf 89}, 166101 (2002).


\bibitem{parry_nature:00}
C. Rasc\'{o}n and  A. O. Parry, Nature {\bf 407}, 986 (2000);
J. Chem. Phys. {\bf 112}, 5157 (2000).



\bibitem{parry_prl:03}
 A. O. Parry,  M. J. Greenaal, and J. M. Romero-Enrique, 
Phys. Rev. Lett. {\bf 90}, 046101 (2003).

\bibitem{abraham_epl:02} 
D. B. Abraham, A. O. Parry, and A. J. Wood, 
Europhys. Lett. {\bf 60}, 106 (2002).

\bibitem{parry_prl:05} 
C. Rasc\'{o}n  and A. O. Parry, 
Phys. Rev. Lett. {\bf 94}, 096103 (2005).

\bibitem{cassie_48}
A. B. D. Cassie, Diss. Faraday Soc. {\bf 3}, 11 (1948).

\bibitem{dietrichbook}
S. Dietrich, in {\it Phase Transitions and Critical Phenomena,} edited by C. Domb and J. L. Lebowitz (Academic, New York, 1988), vol. 12, p. 1.



\bibitem{darbellay}
G. A. Darbellay and J. M. Yeomans,
J.  Phys. A: Math. Gen. {\bf 25}, 4275 (1992).

\bibitem{robbins}
M. O. Robbins, D. Andelman, and J. F. Joanny,
Phys. Rev. A {\bf 43}, 4344 (1991).

\bibitem{goldman} 
S. Goldman, J. Phys. Chem. {\bf 80}, 1697 (1976). 

\bibitem{jasper} J. J. Jasper, J. Phys. Chem. Ref. Data, {\bf 1}, 841 (1972).

\end{thebibliography}
\end{document}